\documentclass[12pt]{iopart}

\usepackage{amssymb}
\usepackage{amsbsy}
\usepackage[dvips]{graphicx}
\usepackage{color}
\usepackage{graphics}

\newcommand{\be}{\begin{equation}} \newcommand{\ee}{\end{equation}}
\newcommand{\bea}{\begin{eqnarray}} \newcommand{\eea}{\end{eqnarray}}
\newcommand{\el}{\nonumber \\}
\newcommand{\re}[1]{(\ref{#1})}
\newcommand{\pat}{\partial}
\newcommand{\adot}{\dot{a}}
\newcommand{\addot}{\ddot{a}}
\newcommand{\rhodot}{\dot{\rho}}

\newcommand{\bx}{\bi{x}}

\newcommand{\brt}[1]{[#1]}
\renewcommand{\H}{\frac{\adot}{a}}
\newcommand{\HH}{\frac{\adot^2}{a^2}}
\newcommand{\sR}{\mathcal{R}}
\newcommand{\sQ}{\mathcal{Q}}
\renewcommand{\a}{\alpha}
\renewcommand{\b}{\beta}

\newcommand{\av}[1]{\langle{#1}\rangle}

\newcommand{\PRD}[1]{{\it Phys. Rev.} {\bf D#1}}
\renewcommand{\PRL}[1]{{\it Phys. Rev. Lett.} {\bf #1}}

\newcommand{\NPB}[1]{{\it Nucl. Phys.} {\bf B#1}}

\newcommand{\MNRAS}[1]{{\it Mon. Not. Roy. Astron. Soc.} {\bf #1}}
\newcommand{\APJ}[1]{{\it Astrophys. J.} {\bf #1}}

\renewcommand{\CQG}[1]{{\it Class. Quant. Grav.} {\bf #1}}
\newcommand{\GRG}[1]{{\it Gen. Rel. Grav.} {\bf #1}}
\renewcommand{\AA}[1]{{\it Astron. \& Astrophys.} {\bf #1}}

\begin{document}
%\baselineskip16pt

%\begin{titlepage}

\title{Constraints on backreaction in dust universes}

\author{Syksy R\"{a}s\"{a}nen}

\address{Rudolf Peierls Centre for Theoretical Physics\\ University of Oxford,
1 Keble Road, Oxford, OX1 3NP, UK}

\ead{syksy.rasanen@iki.fi}

\begin{abstract}

\noindent We study backreaction in dust universes
using exact equations which do not rely on perturbation theory,
concentrating on theoretical and observational constraints.
In particular, we discuss the recent suggestion (in hep-th/0503117)
that superhorizon perturbations could explain present-day
accelerated expansion as a useful example which can be ruled out.
We note that a backreaction explanation of late-time
acceleration will have to involve spatial
curvature and subhorizon perturbations.

\end{abstract}


%\end{titlepage}

\setcounter{secnumdepth}{3}

\section{Introduction}

The idea that averaging inhomogeneous and/or anisotropic
spacetimes can lead to behaviour different
from the homogeneous and isotropic
Friedmann-Robertson-Walker models goes back
to at least 1963 \cite{Shirokov:1963}, and was highlighted
as the 'fitting problem' in 1983 \cite{Ellis}. Under the
name backreaction, the effect of inhomogeneities larger than
the Hubble scale has been studied in general relativity
\cite{Mukhanov, Unruh:1998, Geshnizjani:2002, Brandenberger:2002, Finelli:2003, Geshnizjani:2003, Brandenberger:2004a, Geshnizjani:2004, Brandenberger:2004b, Brandenberger:2004c, Nambu:2005}
and in quantized gravity \cite{Woodard}, particularly
during inflation in the hope that they would provide a graceful exit.
It has also been suggested that backreaction from
inhomogeneities smaller than the Hubble scale could
explain the apparently observed accelerated expansion
of the universe today \cite{Rasanen:2003, Notari:2005},
a modern version of the search for a backreaction solution
to the age problem in the 1990s
\cite{Bildhauer:1991, Russ:1996, Sicka:1999}. For a
useful general formulation of backreaction issues, see
\cite{Buchert:1999a, Buchert:2000, Buchert:2001, Buchert:2003}.
(For a more comprehensive list of backreaction
references, see \cite{Rasanen:2003, Krasinski:1997}.)

Recently it has also been proposed that superhorizon
perturbations could explain the late-time acceleration
\cite{Kolb:2004a, Kolb:2004b, Barausse:2005, Kolb:2005}.
However, in \cite{Geshnizjani:2005} it was claimed that the effect
proposed in \cite{Kolb:2005} amounts only to
a renormalisation of the spatial curvature, and is
thus severely constrained by observations of the
cosmological microwave background and can at any rate
never lead to acceleration. There has also
been more technical criticism, claiming that a proper
accounting of all perturbative terms as well as
more general arguments show that the superhorizon modes
do not lead to acceleration \cite{Flanagan:2005, Hirata:2005}.

We take a viewpoint complementary to that of
\cite{Flanagan:2005, Hirata:2005} and study the
metric presented in \cite{Kolb:2005} as a concrete example
that can used to demonstrate general issues about
backreaction in a dust-dominated universe, relevant
for both super- and subhorizon perturbations. We will
emphasise observational and theoretical constraints and discuss
the relation between backreaction and spatial curvature.

In section 2, we examine backreaction using the formalism of
\cite{Buchert:1999a, Buchert:2001}. We derive
exact bounds on the Hubble parameter and the deceleration
parameter. We find that the metric proposed in \cite{Kolb:2005}
is ruled out on theoretical grounds if it is valid
arbitrarily far into the future, and on observational grounds
even if it is to be valid only up until today.
If the metric were correct, the effect would not reduce
to spatial curvature, though there is a direct link between
spatial curvature and backreaction-driven acceleration.
In section 3, we discuss general lessons for a backreaction
explanation of late-time acceleration, and note that it will
have to involve subhorizon perturbations. We then summarise our
results and point out the caveats.

\section{Constraints on backreaction}

\subsection{The set-up}

\paragraph{The superhorizon backreaction proposal.}

In \cite{Kolb:2005} it is assumed  that the universe has a
homogeneous, isotropic and spatially flat background of
dust, and some perturbations denoted by $\Psi(t,\bx)$
(but no cosmological constant). The
perturbations are divided into two sets of modes,
those with wavelengths short and long compared to the Hubble
scale, $\Psi(t,\bx)=\Psi_s(t,\bx)+\Psi_l(t)$.
From the point of view of an observer seeing a
single Hubble patch, the spatial dependence of
$\Psi_l(t)$ is negligible (by definition).
The metric in the synchronous gauge is given by
\bea \label{metric}
  ds^2 &=& - \rmd t^2 + a_{FRW}(t)^2 e^{-2 \Psi(t,\bx)}  \delta_{ij} \rmd x^i\rmd x^j \el
  &=& - \rmd t^2 + a_{FRW}(t)^2 e^{-2 \Psi_s(t,\bx)-2 \Psi_l(t)}  \delta_{ij} \rmd x^i\rmd x^j \el
  &=& - \rmd t^2 + \left( a_{FRW}(t) e^{-\Psi_l(t)} \right)^2 \delta_{ij} \rmd x^i\rmd x^j \el
  &\equiv& - \rmd t^2 + a(t)^2 \delta_{ij} \rmd x^i\rmd x^j \ ,
\eea

\noindent where $t$ is the proper time measured by a comoving observer,
$a_{FRW}(t)=(t/t_0)^{2/3}$ is the FRW scale factor, the contribution
of the short wavelength modes has been neglected, and on the
last line we have identified the full scale factor as $a(t)$
(our notation is slightly different from that of \cite{Kolb:2005}). The
perturbation is given by $\Psi_l(t)=(a_{FRW}(t)-1)\Psi_{l0}$,
where the normalisation is $a_{FRW}(t_0)=1$ and $\Psi_l(t_0)=0$
today. As a function of proper time, the scale factor is then
\bea \label{a}
  a(t) &=&  e^{\Psi_{l0}} \left( \frac{t}{t_0} \right)^{2/3} e^{-\Psi_{l0}(t/t_0)^{2/3}} \ ,
\eea

\noindent where the normalisation is $a(t_0)=1$.
The metric \re{metric}, \re{a} is claimed in \cite{Kolb:2005}
to be valid to all orders in perturbation theory (a claim
analysed in detail in \cite{Hirata:2005}).
The departure from FRW behaviour is quantified by
$-\Psi_{l0}(t/t_0)^{2/3}$. If $\Psi_{l0}<0$, backreaction
increases the expansion rate and will
eventually lead to acceleration. If $|\Psi_{l0}|\sim1$, the
backreaction is relevant already today. The Hubble parameter
and the deceleration parameter are
\bea
  \label{H} H &\equiv& \H = \frac{2}{3 t} (1-\Psi_{l0}(t/t_0)^{2/3}) \equiv \frac{2}{3 t} (1+x) \\
  \label{q} q &\equiv& - \frac{1}{H^2}\frac{\addot}{a}
  = \frac{1}{2} \frac{ 1 - 3 x - 2 x^2 }{(1 + x)^2} \ ,
\eea

\noindent where a dot indicates derivative with respect to the proper
time $t$ and we have denoted $x\equiv-\Psi_{l0}(t/t_0)^{2/3}$.
From now on, we assume that $\Psi_{l0}<0$, and thus $x>0$.

\paragraph{The criticism regarding spatial curvature.}

In \cite{Geshnizjani:2005}, it was argued that the effect
discussed in \cite{Kolb:2005} amounts to a simple
renormalisation of curvature. The argument is essentially
that one can define a new scale factor such that the metric
looks like a FRW metric with a curvature term.

This argumentation seems to be incorrect, since
for acceleration the relevant issue is the
dependence of the scale factor on the proper time, regardless
of what notation is used for the scale factor (or proper time).
The new scale factor defined in \cite{Geshnizjani:2005}
(essentially \re{a} expanded to first order in $\Psi_{l0}$)
still contains a term proportional to $t^{4/3}$ and thus
implies acceleration. It is true that the metric involves
spatial curvature, but the effect of the perturbations does not
reduce to spatial curvature.
Nevertheless, the relation between the metric given in
\re{metric}, \re{a} and spatial curvature is interesting,
and we will take a closer look at it.

\paragraph{The exact average equations.}

We follow the formalism for analysing the relation between
backreaction and spatial curvature on very general grounds
developed in \cite{Buchert:1999a} for the dust case (and in
\cite{Buchert:2001} for a general ideal fluid).
In the present case with only dust, we assume that the
vorticity is zero, but otherwise allow for general inhomogeneity
and anisotropy. (If vorticity is present, no family of hypersurfaces of
constant proper time exists \cite{Ehlers:1993, Raychaudhuri:1989},
and the formalism of \cite{Buchert:1999a, Buchert:2001} is
inapplicable. See also \cite{Stoeger:1995}.) The Einstein
equation then gives the local equations
\bea
  \label{Rayloc} \dot{\theta} + \frac{1}{3} \theta^2 &=& - \frac{\kappa^2}{2} \rho - 2 \sigma^2 \\
  \label{Hamloc} \frac{1}{3} \theta^2 &=& \kappa^2 \rho - \frac{1}{2} \sR + \sigma^2 \\
  \label{cons} \rhodot + \theta\rho &=& 0 \ ,
\eea

\noindent where a dot still stands for derivative with respect to
the proper time, $\theta(t,\bx)$ is the expansion rate of the local
volume element, $\kappa^2=8\pi G_N$ is the gravitational coupling,
$\rho(t,\bx)$ is the energy density, $\sigma^2(t,\bx)$ is the
shear scalar and $\sR(t,\bx)$ is the spatial curvature.
The energy density and the shear scalar are everywhere non-negative.

Averaging \re{Rayloc}, \re{Hamloc} and \re{cons} over
a spatial domain (here taken to be the observable universe\footnote{The
spatial averaging in \cite{Kolb:2005} is also over a single Hubble patch.})
with volume $V(t)$, we obtain the equations for the average quantities
\bea
  \label{Ray} 3 \frac{\addot}{a} &=& - \frac{\kappa^2}{2}\frac{\langle\rho_0\rangle}{a^3} + \sQ \\
  \label{Ham} 3 \HH &=& \kappa^2\frac{\langle\rho_0\rangle}{a^3} - \frac{1}{2}\av{\sR} - \frac{1}{2}\sQ \ ,
\eea

\noindent where $a$ is the effective scale factor defined
by $a(t)\propto V(t)^{1/3}$, $\rho_0$ is the initial value
of the energy density and $\av{A}$ means the spatial average
of the quantity $A$ over the hypersurface of constant proper time $t$,
\bea
  \av{A} \equiv \frac{ \int d^3 x \sqrt{^{(3)}g} \, A }{ \int d^3 x \sqrt{^{(3)}g} } \ ,
\eea

\noindent where $\sqrt{^{(3)}g}$ is the volume element on the
hypersurface of constant $t$. Note that $V(t)=\int d^3 x \sqrt{^{(3)}g}$,
and $\theta = ( \sqrt{^{(3)}g} )^{-1} \pat_t ( \sqrt{^{(3)}g} )$,
so $\av\theta=3\adot/a\equiv3 H$. The backreaction variable $\sQ$, defined as
\bea \label{Qdef}
  \sQ \equiv \frac{2}{3}\left( \av{\theta^2} - \av{\theta}^2 \right) - 2 \av{\sigma^2} \ ,
\eea

\noindent is a qualitatively new term compared to the FRW equations,
and embodies the effect of inhomogeneity and anisotropy on the
behaviour of the averages.

We emphasise that \re{Ray}, \re{Ham} are exact equations for the
averages. There is no need to assume that any perturbations are small,
or indeed to have any division into background and perturbations.
The equations are general for matter that is a pressureless
ideal fluid with zero vorticity.

Note that \re{Ray}, \re{Ham} have the form of
FRW equations with effective energy density
$\rho_{eff}=\rho-\sQ/(2\kappa^2)$ and effective
pressure $p_{eff}=-\sQ/(2\kappa^2)$ \cite{Buchert:2001},
though the behaviour of $\av{\sR}$ and $\sQ$
is different from the FRW case.
As in the FRW model, the equations \re{Ray} and \re{Ham} are not
independent. The integrability condition is
\bea \label{int}
  \pat_t{\av{\sR}} + 2 \H\av{\sR} = - \dot{\sQ} - 6 \H\sQ \ ,
\eea

\noindent so for $\sQ=0$, the equations reduce to the FRW
case with $\av{\sR}\propto a^{-2}$. Likewise, for $\av{\sR}=0$,
backreaction is reduced to the term $\sQ\propto a^{-6}$.

The system \re{Ray}, \re{Ham} (or either of them together with \re{int})
has two equations for the three unknowns $a, \av{\sR}$ and $\sQ$,
so it is not closed. This means that different inhomogeneous
and/or anisotropic systems sharing the same initial averages can
evolve differently even as far as the averages are concerned, as
could be expected. It also reflects the fact that sources
outside the domain being considered can influence the
evolution within the domain. (In the inflationary scenario
this does not violate causality even when the domain
is the observable universe, since the causally
connected region is much larger.)

Given one of the unknowns $a, \av{\sR}$ and $\sQ$,
the other two are determined. In particular, the
scale factor \re{a} determines the spatial curvature
(and the backreaction variable $\sQ$) uniquely regardless
of whether \re{a} originates from long or short
wavelength perturbations (or whether the inhomogeneity and/or
anisotropy can even be discussed in terms of wave modes).
So, we can take the scale factor \re{a} and plug it
into \re{Ray}, \re{Ham} to see what it implies for the spatial
curvature (assuming the scale factor results from backreaction,
as presented in \cite{Kolb:2005}). However, let us first
discuss how it is possible to have accelerated expansion without
violating the strong energy condition, and obtain theoretical
bounds on the Hubble parameter and acceleration.

\subsection{Theoretical constraints}

\paragraph{General bounds on $H$ and $q$.}

The Raychaudhuri equation \re{Rayloc} shows that
the local acceleration is non-positive at each point in
spacetime,  $\dot{\theta} + \theta^2/3 \leq 0$.
This holds not just for dust, but for any irrotational perfect
fluid satisfying the strong energy condition (but does not
necessarily hold if vorticity is present). However, the
averaged equation \re{Ray} does not rule out the possibility
that the acceleration $\addot$ related to the volume is positive.
The reason for the apparent discrepancy is that the growth
of the volume of the hypersurface of constant proper time
compensates for the decrease in the local expansion rate.
Technically phrased, averaging and taking the time
derivative do not commute:
$\pat_t\av\theta = \av{\dot{\theta}} + \av{\theta^2} - \av{\theta}^2 \geq \av{\dot{\theta}}$.

Though the acceleration is not limited to be negative, it is
possible to derive combined upper limits for the Hubble parameter
and the deceleration parameter. Integrating
$\dot{\theta} + \theta^2/3 \leq 0$, we obtain the bound
\bea \label{thetabound}
  \theta(t,\bx)^{-1} - \theta(t_i,\bx)^{-1} \geq \frac{1}{3} (t-t_i) \ ,
\eea

\noindent which holds at each point in spacetime; $t_i$ is some
initial time. This bound has been derived in \cite{Wald:1984} (and
again in \cite{Nakamura:1995}, where it was applied to the expansion
rate of the universe).
The inequality \re{thetabound} is separately valid in
domains where $\theta(t,\bx)>0$ and in domains where
$\theta(t,\bx)<0$; it is violated if $\theta(t,\bx)$ evolves through zero.
Let's take the initial time to be the big bang time $t_i=0$ and
assume that $\theta(t_i,\bx)$ is positive. We then have the upper bound
$\theta(t,\bx)\leq3/t\equiv\theta_{\textrm{max}}(t)$. This
upper bound is also trivially satisfied in domains where
$\theta(t,\bx)\leq0$, but there is no limit on the absolute value
of $\theta(t,\bx)$, corresponding to the fact that the rate
of collapse is not bounded from below.
Since the average of $\theta$ is at most its maximum value, we
obtain an upper bound on the Hubble parameter in terms of the
age of the universe:
\bea \label{Hbound}
  H = \frac{1}{3} \av{\theta} \leq \frac{1}{3} \theta_{\textrm{max}} = \frac{1}{t} \ .
\eea

The bound \re{Hbound} implies that the acceleration $\addot$
cannot increase (or stay constant, if it is positive) forever:
it must asymptotically go to zero, become negative or
oscillate around zero or a negative value.

We can also put an upper bound on the acceleration as a
function of time and the Hubble parameter.
The averaged Raychaudhuri equation \re{Ray} implies that
\bea \label{qbound}
  q = - \frac{1}{H^2}\frac{\addot}{a} \geq 2 - \frac{2}{9} \frac{\av{\theta^2}}{H^2} \geq 2 - \frac{2}{9} \frac{(\theta_{\textrm{max}})^2}{H^2} = 2 - 2\frac{1}{(H t)^2} \ .
\eea

\noindent In the second inequality we have assumed that
$\av{\theta^2}\leq(\theta_{\textrm{max}})^2$, which
is not necessarily true if $\theta$ is somewhere negative.
This the case in regions of the universe which are
undergoing gravitational collapse. However, the expansion rate
of the universe is generally understood to measure
expansion between regions which have collapsed, so it is
not clear whether the insides of such regions should be
included in the averaging. (According to the inequality
$\dot{\theta} + \theta^2/3 \leq 0$, once a region
starts collapsing, the collapse rate can only increase.
In order for the region to virialise and stabilise,
vorticity or a breakdown in the approximation of treating the
matter as dust \cite{Buchert:2005} is required, so in the present
calculation collapsing regions cannot be treated properly anyway.)

\paragraph{Bounds on the superhorizon proposal.}

The bounds \re{Hbound} and \re{qbound} are valid for any model
where the matter is (irrotational) dust (with a caveat
about the positivity of the expansion rate for the latter bound).
In particular, these bounds can be applied to
the metric presented in \cite{Kolb:2005}. The scale
factor \re{a} is $a\propto x e^{x}$,
where $x=|\Psi_{l0}|(t/t_0)^{2/3}$, so $\addot$
will increase forever and the metric is ruled out.
In more detail, with the expansion rate and the deceleration
parameter given by \re{H} and \re{q}, the theoretical bounds
\re{Hbound} and \re{qbound} reduce to $x\leq1/2$ and
$x\leq(\sqrt{265}-11)/12\approx0.44$, respectively.
Since $x=|\Psi_{l0}|(t/t_0)^{2/3}$ grows without bound, these
bounds will necessarily be violated at some point:
backreaction in a dust universe (without rotation) cannot
produce the scale factor \re{a}, if it is taken to be valid
arbitrarily far into the future.
However, it could still be that the metric is only valid
as an approximation up until today, and the acceleration
would slow down in the future so as to respect the
bounds \re{Hbound}, \re{qbound}.
Assuming the scale factor \re{a} to be valid till today, we get
the bound $|\Psi_{l0}|\leq(\sqrt{265}-11)/12\approx0.44$.
Let us now look at observational constraints
on the metric in light of the evolution given
by \re{Ray}, \re{Ham} and compare to this theoretical bound.

\subsection{Observational constraints}

\paragraph{Evolution of the density parameters.}

We have noted that the local acceleration is everywhere
non-positive, while the acceleration related to the total
volume may be positive. This raises the question of which is
the correct quantity to consider. The answer depends
on what observations one is comparing to.
More generally, there are several different measures of
the expansion rate and acceleration.
For example, number counts (and thus the energy density of dust)
depend on $\theta$ which gives the volume, luminosity
distances depend on different parameters
\cite{Barausse:2005, Flanagan:2005, Hirata:2005}, the rate of
change in the distance $\delta l$ between neighbouring fluid lines
$\dot{\delta l}/\delta l$ is yet another observable given by
$\theta/3 + \sigma_{\a\b}n^\a n^\b$, with 
$\sigma_{\a\b}$ being the shear tensor and $n^\a$ the unit
separation vector between the fluid lines \cite{Raychaudhuri:1989}.
These measures coincide in the homogeneous
and isotropic FRW case, but in general they give different
results for inhomogeneous and/or anisotropic models, so one should
be careful to consider the quantities that are actually being
observed (in the case of acceleration, luminosity distances
of type Ia supernovae). Conversely, agreement between
different measures can be one way to constrain the possibility
that backreaction is responsible for the apparently observed acceleration.

We will not discuss these issues further, but will simply
assume, following \cite{Kolb:2005}, that measures of the
expansion rate and acceleration are related to the scale
factor and its derivatives in the same way as in FRW models.
We will take the scale factor $a(t)\propto V(t)^{1/3}$
to be given by \re{a}, proposed to arise from backreaction
in \cite{Kolb:2005}, and study the observational implications.
As noted earlier, we treat the metric of \cite{Kolb:2005}
as an interesting example with which to look at backreaction.
The predictions of any future backreaction models for $a(t)$
could be analysed in the same manner.
Given the scale factor \re{a}, equations \re{Ray} and \re{Ham}
yield the quantities $\av{\sR}$ and $\sQ$ as
\bea
  \av{\sR} &=& - \frac{2}{3 t^2} \left( 3 - 3 e^{-3 x} + 11 x + 6 x^2 \right) \el
  \sQ &=& \frac{2}{3 t^2} \left( e^{-3 x} - 1 + 3 x + 2 x^2 \right) \ ,
\eea

\noindent where, as before, $x=|\Psi_{l0}|(t/t_0)^{2/3}$.
The spatial curvature is negative, $\av{\sR}<0$, so the universe is open.

The relative contributions of dust, spatial curvature and
the backreaction variable $\sQ$ to the expansion rate
can be discussed in analogy with the FRW case
in terms of the relative densities \cite{Buchert:1999a, Buchert:2003}
\bea
  \Omega_m &\equiv& \frac{\kappa^2\langle\rho_0\rangle a^{-3}}{3 H^2} = \frac{e^{-3x}}{(1+x)^2} \el
  \Omega_{\sR} &\equiv& - \frac{\av{\sR}}{6 H^2} = \frac{1}{4} \frac{ 3 - 3 e^{-3 x} + 11 x + 6 x^2 }{(1+x)^2} \el
  \Omega_\sQ &\equiv& - \frac{\sQ}{6 H^2} = - \frac{1}{4} \frac{ e^{-3 x} - 1 + 3 x + 2 x^2 }{(1+x)^2}  \ ,
\eea

\noindent so that $\Omega_m+\Omega_{\sR}+\Omega_\sQ=1$. As discussed in
\cite{Buchert:1999a, Buchert:2000, Buchert:2003, Buchert:1999b},
backreaction is not important just because $\Omega_\sQ$ can be large,
but also because the presence of a new term changes the behaviour
of the spatial curvature from the FRW case, as shown by \re{int}.

\begin{figure}
\hfill
\begin{minipage}[t]{7cm}
\scalebox{1.11}{\includegraphics[angle=270, width=\textwidth]{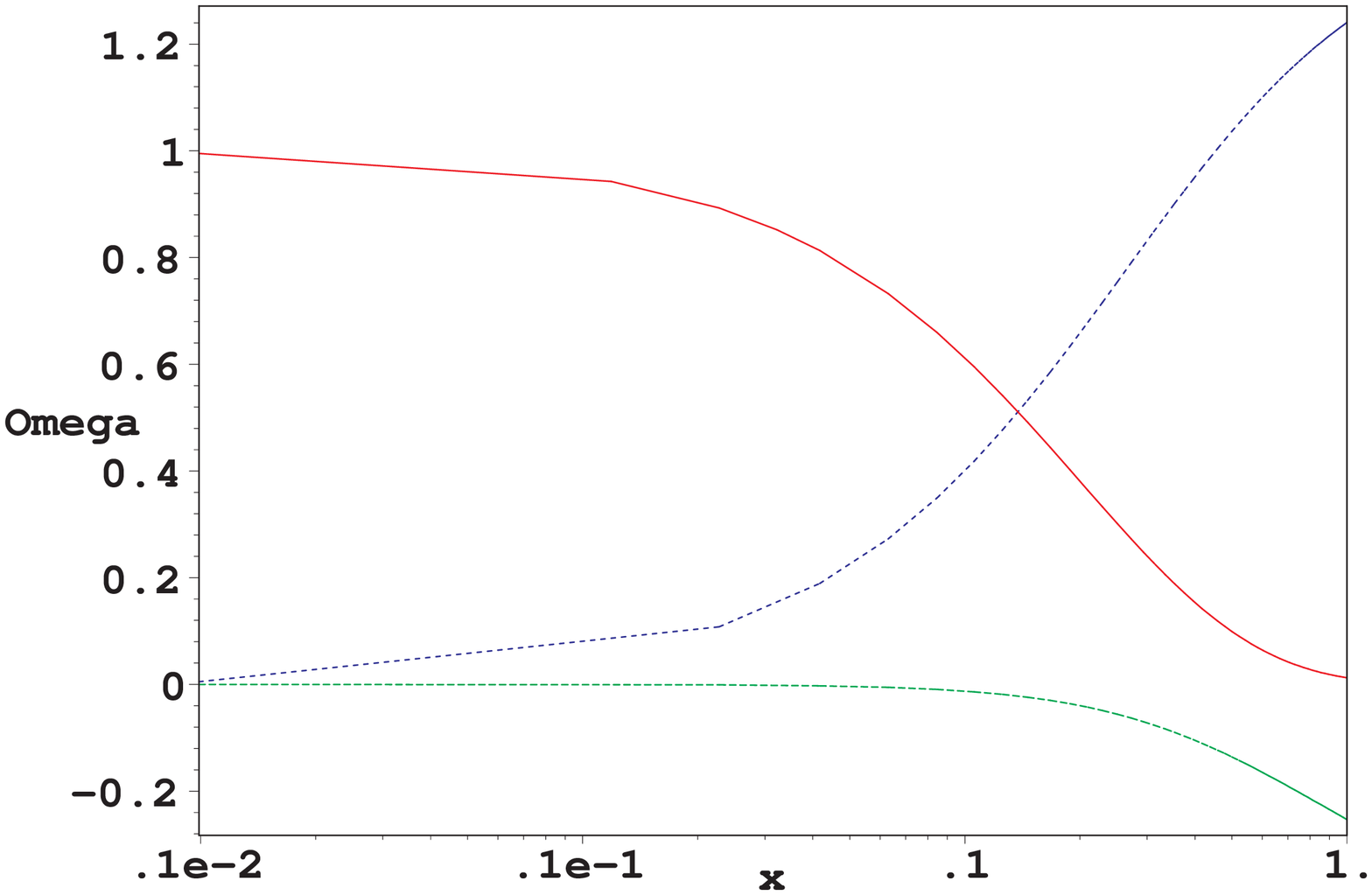}}
\begin{center} {\bf (a)} \end{center}
\end{minipage}
\hfill
\begin{minipage}[t]{7cm}
\scalebox{1.03}{\includegraphics[angle=270, width=\textwidth ]{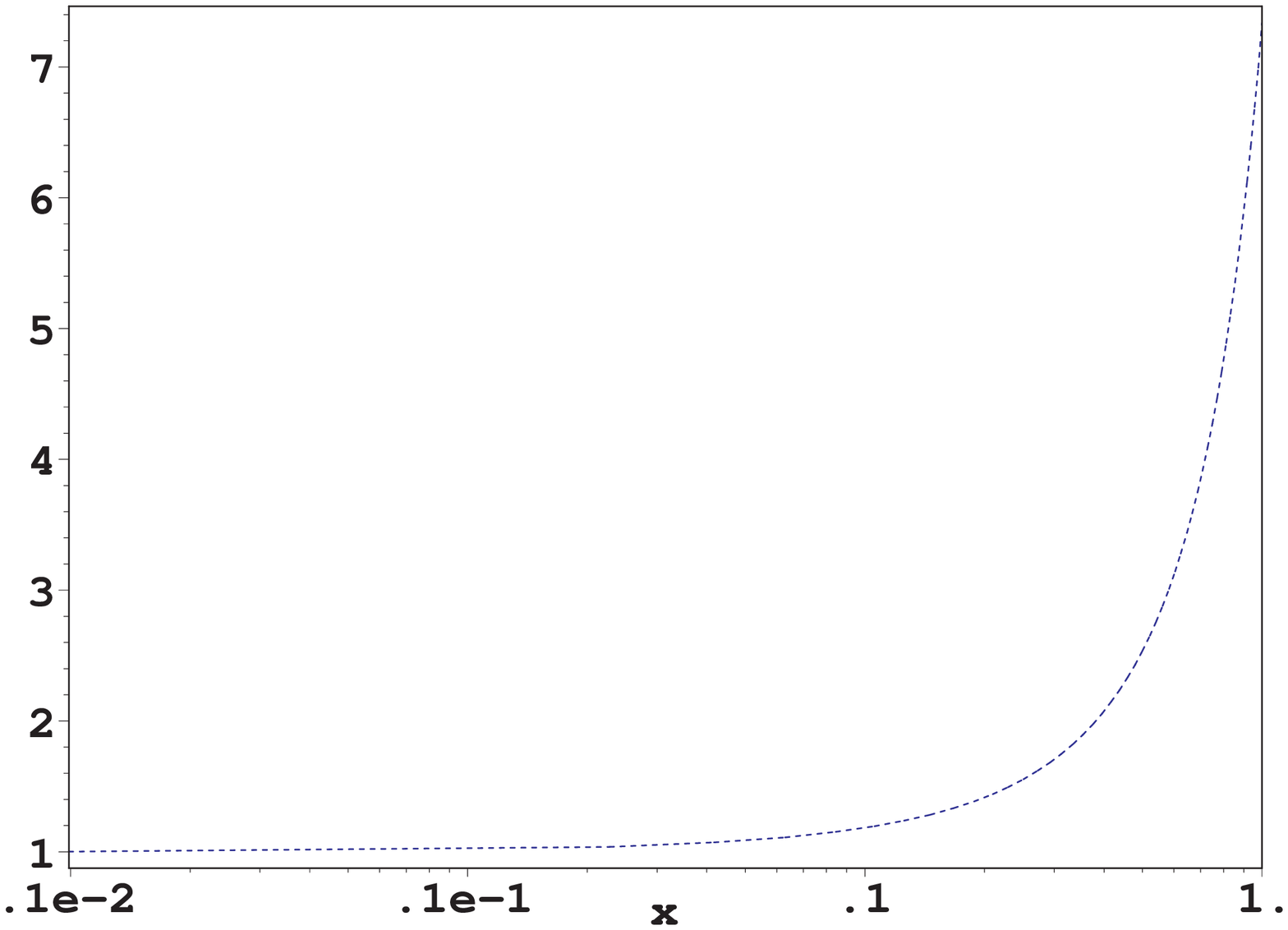}}
\begin{center} {\bf (b)} \end{center}
\end{minipage}
\hfill
\caption{The behaviour of the components
as a function of $x=|\Psi_{l0}|(t/t_0)^{2/3}$.
(a): The relative densities of matter, spatial curvature
and the backreaction variable $\sQ$. Red (solid) is $\Omega_m$,
blue (dash-dot) is $\Omega_{\sR}$ and green (dash) is $\Omega_\sQ$.
(b): The spatial curvature $\av{\sR} a^2$ relative to its
asymptotic past value.}
\end{figure}

The density parameters are plotted in figure 1(a). At early times,
until $x\approx 10^{-2}$, the behaviour is the same
as in the FRW case, with the matter density being dominant
(as is clear from the Hubble law \re{H}). At late times, the
expansion accelerates, the matter density becomes negligible,
and the contributions of the
spatial curvature and the backreaction variable approach
the limiting values $\Omega_{\sR}=3/2$ and $\Omega_\sQ=-1/2$.
Asymptotically, the expansion is driven by the
spatial curvature $\av{\sR}=-9 H^2$ and the backreaction variable
$\sQ=3 H^2$. Note that the behaviour can differ significantly
from the FRW case even when the contribution of the
backreaction variable $\Omega_\sQ$ is numerically small.
For example, for $x=0.1$, we have $\Omega_\sQ\approx-0.01$,
but the contribution of matter has fallen to $\Omega_m\approx0.61$.

\paragraph{Comparing to observations.}

The Hubble parameter, spatial curvature and other 
relevant quantities today are determined by the
single free parameter $|\Psi_{l0}|$ (along with the age of the universe).
We will constrain $|\Psi_{l0}|$ by looking at the observed
values of the Hubble parameter and the deceleration
parameter today, combined with the age of the universe.
From \re{H} and \re{q}, we have
\bea
  \label{H0} H_0 &=& \frac{2}{3 t_0} (1 + |\Psi_{l0}|) \\
  \label{q0} q_0 &=& \frac{1}{2} \frac{ 1 - 3|\Psi_{l0}| - 2 |\Psi_{l0}|^2 }{( 1 + |\Psi_{l0}| )^2} \ .
\eea

Leaving aside the issue that observations might need to be
reanalysed in the context of a backreaction model (as mentioned
above), as well as possible unaccounted systematics in the
determination of the Hubble parameter \cite{Sarkar},
we take $H_0=h$ 100 km/s/Mpc $=72\pm8$ km/s/Mpc, giving $h$
between 0.56 and 0.88 \cite{Freedman:2000}.
The expression \re{H0} can be written as
$1 + |\Psi_{l0}| \approx 2.0 h\times t_0/(13 \textrm{ billion years})$.
Putting the upper limit on $h$ together with the requirement that
the universe is at most 15 billion years old gives
(disregarding correlation in the observational values
of $H_0$ and age) the limit $|\Psi_{l0}| \lesssim 1.0$.
On the other hand, the constraint $q_0\lesssim -0.3$
\cite{Riess:2004} applied to \re{q0} gives
$|\Psi_{l0}|\gtrsim 0.60$. With \re{H0}, this limit can be
written as $h \gtrsim 0.80/(t_0/ $13 billion years)\footnote{This
is with the 'gold' set of SNIae. The 'gold+silver'
set gives $q_0\lesssim -0.4$ leading to $|\Psi_{l0}|\gtrsim0.77$ and
\mbox{$h\gtrsim 0.89/(t_0/ $13 billion years)}. We adopt
0.60 as the upper limit in what follows; the conclusions
would be the same for $|\Psi_{l0}|\gtrsim0.77$.
In \cite{Kolb:2005} it was estimated that $1\gtrsim|\Psi_{l0}|\gtrsim0.25$
gives an acceptable fit to the SNIa data.}.

There is some tension between the age and acceleration constraints:
having enough acceleration requires a large $|\Psi_{l0}|$, which
boosts $H_0$, so $t_0$ needs to be large to bring $H_0$
down in line with the observations. This is related to
the fact that the transition from the FRW dust behaviour
to acceleration is smoother than in $\Lambda$CDM:
the scale factor \re{a} corresponds to the behaviour of a FRW
model with dust replaced by a fluid with the equation of state
$-x/(1+x)$, and two added fluids having the equations of state
$-(1+3x)/(3+3 x)$ and $-(2+3 x)/(3+3 x)$.
The tension is not significant, but we have mentioned it to
show what kind of an impact the age and acceleration constraints
have on backreaction ideas, in particular in relation to the
sharpness of the transition to acceleration.
Obviously, this issue is the same as in models
of dark energy. For a more thorough analysis of
sharp transition into acceleration in light of SNIa and
other cosmological data, see \cite{trans}.

For $|\Psi_{l0}|=0.60$ we have $\Omega_{m0}=0.06$,
$\Omega_{\sR 0}=1.10$ and $\Omega_{\sQ 0}=-0.16$. The universe is
dominated by curvature, with subdominant contributions from
the backreaction variable and matter. Leaving the
small matter density aside, it might seem that the model is
straightforwardly ruled out by constraints on the spatial
curvature from the CMB in models with adiabatic perturbations
(the constraints in the case with correlated adiabatic and
isocurvature perturbations are unknown).

However, when backreaction effects are important
the spatial curvature behaves (at least in a dust universe)
quite differently from the FRW case.
In figure 1(b) we have plotted the quantity $\av{\sR} a^2$, which
is constant in FRW models. The spatial curvature rises faster
than in the FRW case, so it will have less impact in the
past. However, the suppression factor is not large (about 0.32
for $|\Psi_{l0}|=0.60$), and though the observations would need
to be reanalysed with the full inhomogeneous metric, it seems unlikely
that the spatial curvature would be compatible with
the CMB data (at least with adiabatic perturbations).

These observational constraints on the scale factor \re{a}
are independent from the theoretical constraints discussed
earlier. In terms of $h$, the  theoretical
bounds \re{Hbound}, \re{qbound} give
\bea
  \label{h2} h \times t_0 / (13 \textrm{ billion years}) \lesssim 0.75 \\
  \label{q2} q_0 \gtrsim 2 - \frac{1.13}{(h \times t_0 / (13 \textrm{ billion years}))^2} \ .
\eea

The limit $h \times t_0 / (13 \textrm{ billion years})\gtrsim0.80$
that we obtained previously by comparing the scale factor \re{a}
to observations of $q_0$ is in contradiction with \re{h2}.
Applied to \re{q2}, the limit on $h$ implies $q_0\gtrsim0.23$, which is
even more clearly inconsistent. So, the observational lower limit
$|\Psi_{l0}|\gtrsim0.60$ is strongly contradicted by the
theoretical upper bound $|\Psi_{l0}|\lesssim0.44$.
This means that backreaction cannot produce the scale factor
\re{a}, even as an approximation valid only up until today.
In physical terms, the slow onset of acceleration in \re{a}
means that to match the observations, the universe has to be far
in the regime where backreaction is important, which in turn
makes the Hubble parameter higher than irrotational inhomogeneity
or anisotropy in dust can produce.

The bounds \re{h2}, \re{q2} can be used together with observations
to constrain backreaction ideas
in general. In practice, \re{q2} gives the more stringent
constraint: the limit $q_0\lesssim-0.3$ gives
$h \times t_0 / (13 \textrm{ billion years})\lesssim0.70$.
This bound is below the current observational
central value: there is no contradiction at present, but
future observations may be able to show that the bound
is violated. This would mean that a backreaction
explanation for the observed acceleration is ruled out (with
caveats about vorticity, positivity of the expansion rate
and the interpretation of observations).

\section{Discussion}

\paragraph{Lessons for backreaction.}

The analysis of the metric \re{metric}, \re{a} has
allowed us to look at general features of backreaction in a
dust-dominated universe (some brought up already in
\cite{Buchert:1999a, Buchert:2000, Buchert:1999b})
in a quantitative manner.

The backreaction variable $\sQ$
contributes positively to the acceleration \re{Ray}, but
negatively to the square of the Hubble parameter \re{Ham}.
If backreaction increases the expansion rate, this
can only be consistent if there is a compensating
spatial curvature so that $|\av{\sR}|>\sQ$ and $\av{\sR}<0$.
Significant backreaction implies significant spatial curvature.
The fact that spatial curvature is integral to
backreaction\footnote{At least with dust: the case with
non-zero pressure is more complicated \cite{Buchert:2001}.} does not
mean that backreaction as a source of late-time acceleration is
ruled out, since the behaviour of spatial curvature is different
from the FRW case. The CMB data would need to be analysed with
the particular inhomogeneous metric leading to backreaction, not
just the averaged equations. (Note that since the spatial curvature
is time-dependent, perturbations cannot be expanded in terms
of Fourier modes or the basis functions for other spaces
of constant curvature.)
But it seems likely that in order for the spatial curvature not to
conflict with observations, it should decrease rapidly
towards the past.
(Note that negative spatial curvature boosts the amplitudes
of the low multipoles of the CMB via the Integrated Sachs-Wolfe
effect. In general, spatial curvature leads to a stronger ISW
effect than a cosmological constant with the same relative density,
since spatial curvature is important for a longer period of time,
so a transition that is more rapid than in the
$\Lambda$CDM model could yield a smaller ISW effect.)

In other words, in order for
a backreaction explanation to be viable, the transition from
standard FRW behaviour to acceleration would need to be sharper
than in the scale factor \re{a}. This is also suggested by the
tension between the observational age and acceleration constraints.
A sharp transition would sit well with the idea of backreaction
from small wavelength modes related to structure formation: the
expansion rate would be the FRW one until structure formation
becomes important, and then change rapidly.
(Backreaction from structure formation would also present a
solution to the coincidence problem, unlike
backreaction from superhorizon modes.)

Note that according to \re{Ray} and \re{Qdef} accelerated
expansion requires a variance of the expansion rate of
order unity\footnote{Note that this is the variance within
a single Hubble patch, not between different Hubble patches as in
\cite{Kolb:2004a, Kolb:2004b, Barausse:2005, Kolb:2005}.}.
Though we have simply assumed that superhorizon
modes lead to the scale factor \re{a},  it in
fact implies significant subhorizon perturbations
(if it arises from backreaction). This shows that, observational
constraints aside, superhorizon perturbations
alone cannot lead to late-time acceleration.

The relation between variance
and backreaction is in agreement with the suggestion made
in \cite{Rasanen:2003} (and borne out in \cite{Rasanen:2004b}
for the Lema\^{\i}tre-Tolman-Bondi model) that the Weyl tensor could
provide a measure of backreaction. On the other hand, large
backreaction from highly isotropic long-wavelength modes
as in \cite{Kolb:2005} (which presumably do not contribute
much to the Weyl tensor) would contradict this idea.

Let us also note that the isotropy of the CMB would
naively seem to rule out variations larger than $10^{-5}$
in the expansion rate in different directions\footnote{I am
grateful to David Lyth for this point.}, and it seems unclear
how this could be reconciled with the large variance needed
for backreaction to produce acceleration. However, the link
between geometric inhomogeneity and/or anisotropy
and the isotropy of the CMB is not
straightforward, and it is not clear that such a large
variance is ruled out either. For example, it is possible
for a spacetime to be very anisotropic even though the CMB
looks almost isotropic \cite{Wainwright}.
See \cite{Stoeger:1995, Maartens} for a careful analysis of
the limits the isotropy of the CMB places on the spatial
variation of the expansion rate and related quantities, and
\cite{Clarkson} for further discussion and counter-examples\footnote{Note
that one of the key assumptions of \cite{Stoeger:1995} is that
$\theta>0$ everywhere.}.
For another point of view, note that the difference in the expansion
rate between expanding regions and regions which have broken away
from the expansion and collapsed is indeed of order one, without
contradicting the isotropy of the CMB.

A possible way of avoiding the large variance would be
that backreaction does not lead to acceleration
as measured by the volume expansion $\theta$, but does
give real or apparent acceleration according to
other measures, for example the luminosity
distance (along the lines of \cite{Barausse:2005}).
In fact, it has been claimed that
cluster counts (which depend on $\theta$ more
directly than SNIa luminosity distances)
are consistent with deceleration and
inconsistent with acceleration \cite{Blanchard}.

\paragraph{Conclusion.}

We have looked at constraints on backreaction-driven
acceleration in a dust-dominated universe,  applying
the exact formalism for backreaction developed in
\cite{Buchert:1999a}. In particular, we have analysed
the model for late-time acceleration presented in
\cite{Kolb:2005} as a useful example, with emphasis on
theoretical and observational constraints as well as spatial
curvature, which was claimed in \cite{Geshnizjani:2005} to
explain away the effect.

We have derived exact bounds for the expansion rate and acceleration,
valid for any models where the matter is irrotational dust.
In particular, acceleration cannot increase forever. This rules
out the metric proposed in \cite{Kolb:2005} on theoretical grounds,
if it is to be valid arbitrarily far into the future.
We also have also shown that, when supplemented with observational
bounds on the Hubble parameter or the deceleration parameter,
the theoretical bounds rule out the metric proposed in
\cite{Kolb:2005} even as an approximation valid only up until today.

If the scale factor in \cite{Kolb:2005} were correct,
we find that the backreaction would not reduce to spatial
curvature, and the universe would accelerate. However, growth
of spatial curvature is an integral part of backreaction
leading to acceleration, and the slow transition
into acceleration in \cite{Kolb:2005} implies that
curvature is important for a long period, which is
probably not compatible with the CMB data.
The same conclusion applies to any model aiming
to explain late-time acceleration with backreaction, whether
from superhorizon modes or inhomogeneities related to
structure formation. This suggests that for backreaction
to be a possible explanation, the transition from standard
FRW behaviour to acceleration should be rapid.
Acceleration from backreaction also implies
that the variance of the expansion rate within the observable
universe is large, showing that superhorizon perturbations
alone cannot be responsible (which is the case in \cite{Kolb:2005}).

The analysis does not depend on any assumptions regarding the
amplitude and wavelength of the inhomogeneities and/or
anisotropies. However, it is assumed that the
cosmic fluid is completely pressureless, ideal (for example,
has no viscosity) and irrotational. Possibilities
for avoiding the growth of spatial curvature thus include the
presence of vorticity (as mentioned in \cite{Hirata:2005}),
or a breakdown in the approximation of treating the cosmic
matter as a pressureless ideal fluid, as conjectured in
\cite{Schwarz:2002} (see also \cite{Buchert:2005}).
Another possibility is that instead of giving acceleration,
backreaction would change the interpretation of
observations so that they are consistent with decelerating expansion.

The above arguments do not apply to backreaction during inflation
in the early universe driven by the cosmological constant or
a scalar field. However, a single scalar field can be
described in terms of an ideal fluid, and
it would be interesting to extend the analysis to
backreaction in single field inflation, as suggested in
\cite{Buchert:2001}.

Finally, let us note that even though the proposal in \cite{Kolb:2005}
of backreaction from superhorizon modes is not
correct, it provides a useful example of a backreaction model which it
is possible to analytically discuss. It is more difficult
to calculate the backreaction of small wavelength modes, but it would
be helpful to find some, perhaps simplified, model to concretely
demonstrate the general issues discussed here in the context of
subhorizon perturbations.

\ack

I thank David Lyth and participants at
CosmoCoffee (http://cosmocoffee.info/) for useful discussions.
The research has been supported by PPARC grant PPA/G/O/2002/00479.\\

\appendix

\setcounter{section}{1}


\begin{thebibliography}{99}

\bibitem{Shirokov:1963} Shirokov M F and Fisher I Z,
\newblock {\it Isotropic Space with Discrete Gravitational-Field Sources. On the Theory of a Nonhomogeneous Universe}, 1963
\newblock {\it Sov. Astron. J.} {\bf 6} 699
\newblock Reprinted in \GRG{30} 1411, 1998
%%CITATION = NONE;%%

\bibitem{Ellis} Ellis G F R,
\newblock {\it Relativistic cosmology: its nature, aims and problems}, 1983
\newblock The invited papers of the 10th international conference on general relativity and gravitation, p 215
%%CITATION = NONE;%%
\nonum Ellis G F R and Stoeger W,
\newblock {\it The 'fitting problem' in cosmology}, 1987
\newblock \CQG{4} 1697
%%CITATION = NONE;%%

\bibitem{Mukhanov} Mukhanov V F, Abramo L R W and Brandenberger R H,
\newblock {\it Back Reaction Problem for Gravitational Perturbations}, 1997
\newblock \PRL{78} 1624
\newblock \brt{gr-qc/9609026}
%%CITATION = GR-QC 9609026;%%
\nonum Abramo L R W, Brandenberger R H and Mukhanov V F,
\newblock {\it The Energy-Momentum Tensor for Cosmological Perturbations}, 1997
\newblock \PRD{56} 3248
\newblock \brt{gr-qc/9704037}
%%CITATION = GR-QC 9704037;%%

\bibitem{Unruh:1998} Unruh W,
\newblock {\it Cosmological long wavelength perturbations}
\newblock \brt{astro-ph/9802323}
%%CITATION = ASTRO-PH 9802323;%%

\bibitem{Geshnizjani:2002} Geshnizjani G and Brandenberger R,
\newblock {\it Back Reaction And Local Cosmological Expansion Rate}, 2002
\newblock \PRD{66} 123507
\newblock \brt{gr-qc/0204074}
%%CITATION = GR-QC 0204074;%%

\bibitem{Brandenberger:2002} Brandenberger R H,
\newblock {\it Back Reaction of Cosmological Perturbations and the Cosmological Constant Problem}
\newblock \brt{hep-th/0210165}
%%CITATION = HEP-TH 0210165;%%

\bibitem{Finelli:2003} Finelli F, Marozzi G, Vacca G P and Venturi G,
\newblock {\it Energy-Momentum Tensor of Cosmological Fluctuations during Inflation}, 2004
\newblock \PRD{69} 123508
\newblock \brt{gr-qc/0310086}
%%CITATION = GR-QC 0310086;%%

\bibitem{Geshnizjani:2003} Geshnizjani G and Brandenberger R,
\newblock {\it Back Reaction Of Perturbations In Two Scalar Field Inflationary Models}, 2005
\newblock JCAP0504(2005)006
\newblock \brt{hep-th/0310265}
%%CITATION = HEP-TH 0310265;%%

\bibitem{Brandenberger:2004a} Brandenberger R and Mazumdar A,
\newblock {\it Dynamical Relaxation of the Cosmological Constant and Matter Creation in the Universe}, 2004
\newblock JCAP0408(2004)015
\newblock \brt{hep-th/0402205}
%%CITATION = HEP-TH 0402205;%%

\bibitem{Geshnizjani:2004} Geshnizjani G and Afshordi N,
\newblock {\it Coarse-Grained Back Reaction in Single Scalar Field Driven Inflation}, 2005
\newblock JCAP0501(2005)011
\newblock \brt{gr-qc/0405117}
%%CITATION = GR-QC 0405117;%%

\bibitem{Brandenberger:2004b} Brandenberger R H and Lam C S,
\newblock {\it Back-Reaction of Cosmological Perturbations in the Infinite Wavelength Approximation}
\newblock \brt{hep-th/0407048}
%%CITATION = HEP-TH 0407048;%%

\bibitem{Brandenberger:2004c} Brandenberger R H and Martin J,
\newblock {\it Back-reaction and the trans-Planckian problem of inflation revisited}, 2005
\newblock \PRD{71} 023504
\newblock \brt{hep-th/0410223}
%%CITATION = HEP-TH 0410223;%%

\bibitem{Nambu:2005} Nambu Y,
\newblock {\it The separate universe and the back reaction of long wavelength fluctuations}, 2005
\newblock \PRD{71} 084016
\newblock \brt{gr-qc/0503111}
%%CITATION = GR-QC 0503111;%%

\bibitem{Woodard} Tsamis N C and Woodard R P,
\newblock {\it Quantum Gravity Slows Inflation}, 1996
\newblock \NPB{474} 235
\newblock \brt{hep-ph/9602315}
%%CITATION = HEP-PH 9602315;%%
\nonum Tsamis N C and Woodard R P,
\newblock {\it The Quantum Gravitational Back-Reaction on Inflation}, 1997
\newblock {\it Annals Phys.} {\bf 253} 1
\newblock \brt{hep-ph/9602316}
%%CITATION = HEP-PH 9602316;%%
\nonum Abramo L R, Tsamis N C and Woodard R P,
\newblock {\it Cosmological Density Perturbations From A Quantum Gravitational Model Of Inflation}, 1999
\newblock {\it Fortsch. Phys.} {\bf 47} 389
\newblock \brt{astro-ph/9803172}
%%CITATION = ASTRO-PH 9803172;%%
\nonum Woodard R P,
\newblock {\it Effective Field Equations of the Quantum Gravitational Back-Reaction on Inflation}
\newblock \brt{astro-ph/0111462}
%%CITATION = ASTRO-PH 0111462;%%
\nonum Woodard R P,
\newblock {\it A leading logarithm approximation for inflationary quantum field theory},
\newblock \brt{astro-ph/0502556}
%%CITATION = ASTRO-PH 0502556;%%

\bibitem{Rasanen:2003} R\"{a}s\"{a}nen S,
\newblock {\it Dark energy from backreaction}, 2004
\newblock JCAP0402(2004)003
\newblock \brt{astro-ph/0311257}
%%CITATION = ASTRO-PH 0311257;%%

\nonum R\"{a}s\"{a}nen S,
\newblock {\it Backreaction of linear perturbations and dark energy}
\newblock \brt{astro-ph/0407317}
%%CITATION = ASTRO-PH 0407317;%%

\bibitem{Notari:2005} Notari A,
\newblock {\it Late time failure of Friedmann equation}
\newblock \brt{astro-ph/0503715}
%%CITATION = ASTRO-PH 0503715;%%

\bibitem{Bildhauer:1991} Bildhauer S and Futamase T,
\newblock{\it The Age Problem in Inhomogeneous Universes}, 1991
\newblock \GRG{23} 1251
%%CITATION = NONE;%%

\bibitem{Russ:1996} Russ H, Soffel M H, Kasai M and B\"{o}rner G,
\newblock {\it Age of the Universe: Influence of the Inhomogeneities on the global Expansion-Factor}, 1997
\newblock \PRD{56} 2044
\newblock \brt{astro-ph/9612218}
%%CITATION = ASTRO-PH 9612218;%%

\bibitem{Sicka:1999} Sicka C, Buchert T and Kerscher M,
\newblock {\it Backreaction in cosmological models}
\newblock \brt{astro-ph/9907137}
%%CITATION = ASTRO-PH 9907137;%%

\bibitem{Buchert:1999a} Buchert T,
\newblock {\it On average properties of inhomogeneous fluids in general relativity I: dust cosmologies}, 2000
\newblock \GRG{32} 105
\newblock \brt{gr-qc/9906015}
%%CITATION = GR-QC 9906015;%%

\bibitem{Buchert:2000} Buchert T,
\newblock {\it On average properties of inhomogeneous cosmologies}, 2000
\newblock Proceedings of the 9th JGRG meeting, p 306
\newblock \brt{gr-qc/0001056}
%%CITATION = GR-QC 0001056;%%

\bibitem{Buchert:2001} Buchert T,
\newblock {\it On average properties of inhomogeneous fluids in general relativity II: perfect fluid cosmologies}, 2001
\newblock \GRG{33} 1381
\newblock \brt{gr-qc/0102049}
%%CITATION = GR-QC 0102049;%%

\bibitem{Buchert:2003} Buchert T and Carfora M,
\newblock {\it The Cosmic Quartet - Cosmological Parameters of a Smoothed Inhomogeneous Spacetime}
\newblock \brt{astro-ph/0312621}
%%CITATION = ASTRO-PH 0312621;%%

\bibitem{Krasinski:1997} Krasi\'{n}ski A,
\newblock {\it Inhomogeneous Cosmological Models}, 1997
\newblock Cambridge University Press, Cambridge
%%CITATION = NONE;%%

\bibitem{Kolb:2004a} Kolb E W, Matarrese S, Notari A and Riotto A,
\newblock {\it The effect of inhomogeneities on the expansion rate of the universe}, 2005
\newblock \PRD{71} 023524
\newblock \brt{hep-ph/0409038}
%%CITATION = HEP-PH 0409038;%%

\bibitem{Kolb:2004b} Kolb E W, Matarrese S, Notari A and Riotto A,
\newblock {\it Cosmological influence of super-Hubble perturbations}, 2005
\newblock {\it Mod. Phys. Lett.} {\bf A20} 2705
\newblock \brt{astro-ph/0410541}
%%CITATION = ASTRO-PH 0410541;%%

\bibitem{Barausse:2005} Barausse E, Matarrese S and Riotto A,
\newblock {\it The Effect of Inhomogeneities on the Luminosity Distance-Redshift Relation: is Dark Energy Necessary in a Perturbed Universe?}, 2005
\newblock \PRD{71} 063537
\newblock \brt{astro-ph/0501152}
%%CITATION = ASTRO-PH 0501152;%%

\bibitem{Kolb:2005} Kolb E W, Matarrese S, Notari A and Riotto A,
\newblock {\it Primordial inflation explains why the universe is accelerating today}
\newblock \brt{hep-th/0503117}
%%CITATION = HEP-TH 0503117;%%

\bibitem{Geshnizjani:2005} Geshnizjani G, Chung D J H and Afshordi N,
\newblock {\it Do Large-Scale Inhomogeneities Explain Away Dark Energy?}, 2005
\newblock \PRD{72} 023517
\newblock \brt{astro-ph/0503553}
%%CITATION = ASTRO-PH 0503553;%%

\bibitem{Flanagan:2005} Flanagan E E,
\newblock {\it Can superhorizon perturbations drive the acceleration of the Universe?}, 2005
\newblock \PRD{71} 103521
\newblock \brt{hep-th/0503202}
%%CITATION = HEP-TH 0503202;%%

\bibitem{Hirata:2005} Hirata C M and Seljak U,
\newblock {\it Can superhorizon cosmological perturbations explain the acceleration of the universe?}, 2005
\newblock \PRD{72} 083501
\newblock \brt{astro-ph/0503582}
%%CITATION = ASTRO-PH 0503582;%%

\bibitem{Ehlers:1993} Ehlers J,
\newblock {\it Contributions to the relativistic mechanics of continuous media}, 1993
\newblock \GRG{25} 1225
\newblock (Originally 1961, Abh.\  Akad.\  Wiss.\  Lit.\  Mainz.\  Nat.\  Kl.\  {\bf 11} 793)
%%CITATION = GRGVA,25,1225;%%

\bibitem{Raychaudhuri:1989} Raychaudhuri A K,
\newblock {\it An Approach To Anisotropic Cosmologies}, 1989
\newblock Gravitation, gauge theories and the early universe, ed Iyer B R \etal
\newblock Kluwer, Dordrecht
\newblock p 89
%\href{http://www.slac.stanford.edu/spires/find/hep/www?irn=2147238}{SPIRES entry}
%%CITATION = NONE;%%

\bibitem{Stoeger:1995} Stoeger W R, Maartens R and Ellis G F R,
\newblock {\it Proving almost homogeneity of the universe: An Almost Ehlers-Geren-Sachs theorem}, 1995
\newblock \APJ{443} 1
%%CITATION = ASJOA,443,1;%%

\bibitem{Wald:1984} Wald R M,
\newblock {\it General Relativity}, 1984
\newblock The University of Chicago Press, Chicago
\newblock p 220
%%CITATION = NONE;%%

\bibitem{Nakamura:1995} Nakamura T, Nakao K-i, Chiba T and Shiromizu T,
\newblock {\it Volume expansion rate and age of universe}, 1995
\newblock \MNRAS{276} L41
\newblock \brt{astro-ph/9507085}
%%CITATION = ASTRO-PH 9507085;%%

\bibitem{Buchert:2005} Buchert T and Dom\'\i nguez A,
\newblock {\it Adhesive Gravitational Clustering}, 2005
\newblock \AA{438} 443
\newblock \brt{astro-ph/0502318}
%%CITATION = ASTRO-PH 0502318;%%

\bibitem{Buchert:1999b} Buchert T, Kerscher M and Sicka C,
\newblock {\it Backreaction of inhomogeneities on the expansion: the evolution of cosmological parameters}, 2000
\newblock \PRD{62} 043525
\newblock \brt{astro-ph/9912347}
%%CITATION = ASTRO-PH 9912347;%%

\bibitem{Sarkar} Blanchard A, Douspis M, Rowan-Robinson M and Sarkar S,
\newblock {\it An alternative to the cosmological 'concordance model'}, 2003
\newblock \AA{412} 35
\newblock \brt{astro-ph/0304237}
%%CITATION = ASTRO-PH 0304237;%%

\nonum Sarkar S,
\newblock {\it Measuring the cosmological density perturbation}, 2005
\newblock {\it Nucl. Phys.} {\bf B} {\it Proc. Suppl.} 148 1
\newblock \brt{hep-ph/0503271}
%%CITATION = HEP-PH 0503271;%%

\bibitem{Freedman:2000} Freedman W L \etal,
\newblock {\it Final Results from the Hubble Space Telescope Key Project to Measure the Hubble Constant}, 2001
\newblock \APJ{553} 47
\newblock \brt{astro-ph/0012376}
%%CITATION = ASTRO-PH 0012376;%%

\bibitem{Riess:2004} Riess A G \etal (Supernova Search Team Collaboration),
\newblock {\it Type Ia Supernova Discoveries at $z>1$ From the Hubble Space Telescope: Evidence for Past Deceleration and Constraints on Dark Energy Evolution}, 2004
\newblock \APJ{607} 665
\newblock \brt{astro-ph/0402512}
%%CITATION = ASTRO-PH 0402512;%%

\bibitem{trans} Bassett B A, Kunz M, Silk J and Ungarelli C,
\newblock {\it A late-time transition in the cosmic dark energy?}, 2002
\newblock \MNRAS{336} 1217
\newblock \brt{astro-ph/0203383}
%%CITATION = ASTRO-PH 0203383;%%

\nonum Alam U, Sahni V and Starobinsky A A,
\newblock{\it Is there Supernova Evidence for Dark Energy Metamorphosis?}, 2004
\newblock \MNRAS{354} 275
\newblock \brt{astro-ph/0311364}
%%CITATION = ASTRO-PH 0311364;%%

\nonum Alam U, Sahni V and Starobinsky A A,
\newblock{\it The case for dynamical dark energy revisited}, 2004
\newblock JCAP0406(2004)008
\newblock \brt{astro-ph/0403687}
%%CITATION = ASTRO-PH 0403687;%%

\nonum Jonsson J, Goobar A, Amanullah R and Bergstrom L,
\newblock{\it No evidence for Dark Energy Metamorphosis?}, 2004
\newblock JCAP09(2004)007
\newblock \brt{astro-ph/0404468}
%%CITATION = ASTRO-PH 0404468;%%

\nonum Corasaniti P S, Kunz M, Parkinson D, Copeland E J and Bassett B A,
\newblock{\it The foundations of observing dark energy dynamics with the Wilkinson Microwave Anisotropy Probe}, 2004
\newblock \PRD{70} 083006
\newblock \brt{astro-ph/0406608}
%%CITATION = ASTRO-PH 0406608;%%

\nonum Alam U, Sahni V and Starobinsky A A,
\newblock{\it Rejoinder to No Evidence of Dark Energy Metamorphosis}
\newblock \brt{astro-ph/0406672}
%%CITATION = ASTRO-PH 0406672;%%

\nonum Hannestad S and M\"{o}rtsell E,
\newblock{\it Cosmological constraints on the dark energy equation of state and its evolution}, 2004
\newblock JCAP0409(2004)001 
\newblock \brt{astro-ph/0407259}
%%CITATION = ASTRO-PH 0407259;%%

\nonum Bassett B A, Corasaniti P S and Kunz M,
\newblock{\it The essence of quintessence and the cost of compression}, 2004
\newblock \APJ{617} L1
\newblock \brt{astro-ph/0407364}
%%CITATION = ASTRO-PH 0407364;%%

\bibitem{Rasanen:2004b} R\"{a}s\"{a}nen S,
\newblock {\it Backreaction in the Lema\^{\i}tre-Tolman-Bondi model}, 2004
\newblock JCAP0411(2004)010
\newblock \brt{gr-qc/0408097}
%%CITATION = GR-QC 0408097;%%

\bibitem{Wainwright} Wainwright J, Hancock M J and Uggla C,
\newblock {\it Asymptotic self-similarity breaking at late times in cosmology}, 1999
\newblock \CQG{16} 2577
\newblock \brt{gr-qc/9812010}
%%CITATION = GR-QC 9812010;%%

\nonum Nilsson U S, Uggla C, Wainwright J and Lim W C,
\newblock {\it An almost isotropic cosmic microwave temperature does not imply an almost isotropic universe}, 1999
\newblock \APJ{522} L1
\newblock \brt{astro-ph/9904252}
%%CITATION = ASTRO-PH 9904252;%%

\nonum Lim W C, Nilsson U S and Wainwright J,
\newblock {\it Anisotropic universes with isotropic cosmic microwave background radiation}, 2001
\newblock \CQG{18} 5583
\newblock \brt{gr-qc/9912001}
%%CITATION = GR-QC 9912001;%%

\bibitem{Maartens} Maartens R, Ellis G F R and Stoeger W R,
\newblock {\it Limits on anisotropy and inhomogeneity from the cosmic background radiation}, 1995
\newblock \PRD{51} 1525
\newblock \brt{astro-ph/9501016}
%%CITATION = ASTRO-PH 9501016;%%

\nonum Maartens R, Ellis G F R and Stoeger W R,
\newblock {\it Improved limits on anisotropy and inhomogeneity from the cosmic background radiation}, 1995
\newblock \PRD{51} 5942
%%CITATION = NONE;%%

\nonum Maartens R, Ellis G F R and Stoeger W R,
\newblock {\it Anisotropy and inhomogeneity of the universe from $\Delta T/T$}, 1996
\newblock \AA{309} L7
\newblock \brt{astro-ph/9510126}
%%CITATION = ASTRO-PH 9510126;%%

\nonum Stoeger W R, Araujo M and Gebbie T,
\newblock {\it The Limits on Cosmological Anisotropies and Inhomogeneities from COBE Data}, 1997
\newblock \APJ{476} 435
\newblock \brt{astro-ph/9904346}
%%CITATION = ASTRO-PH 9904346;%%

\bibitem{Clarkson} Clarkson C A and Barrett R K,
\newblock {\it Does the Isotropy of the CMB Imply a Homogeneous Universe? Some Generalised EGS Theorems}, 1999
\newblock \CQG{16} 3781
\newblock \brt{gr-qc/9906097}
%%CITATION = GR-QC 9906097;%%

\nonum Barrett R K and Clarkson C A,
\newblock {\it Undermining the Cosmological Principle: Almost Isotropic Observations in Inhomogeneous Cosmologies}, 2000
\newblock \CQG{17} 5047
\newblock \brt{astro-ph/9911235}
%%CITATION = ASTRO-PH 9911235;%%

\nonum Clarkson C A, Coley A A, O'Neill E S D, Sussman R A and Barrett R K,
\newblock {\it Inhomogeneous cosmologies, the Copernican principle and the cosmic microwave background: More on the EGS theorem}, 2003
\newblock \GRG{35} 969
\newblock \brt{gr-qc/0302068}
%%CITATION = GR-QC 0302068;%%

\bibitem{Blanchard} Vauclair S C \etal ,
\newblock {\it The XMM--NEWTON Omega Project: II. Cosmological implications from the high redshift L-T relation of X-ray clusters}, 2003
\newblock \AA{412} L37
\newblock \brt{astro-ph/0311381}
%%CITATION = ASTRO-PH 0311381;%%

\nonum Blanchard A,
\newblock {\it Cosmological Interpretation from High Redshift Clusters Observed Within the XMM-Newton $\Omega$-Project}
\newblock \brt{astro-ph/0502220}
%%CITATION = ASTRO-PH 0502220;%%

\bibitem{Schwarz:2002} Schwarz D J,
\newblock {\it Accelerated expansion without dark energy},
\newblock \brt{astro-ph/0209584}
%%CITATION = ASTRO-PH 0209584;%%

\end{thebibliography}
\end{document}